\documentclass[twocolumn,separability,Amsterdam,nobibnotes, aps, pra,showpacs,preprintnumbers]{revtex4-1}
\usepackage{amssymb}
\usepackage{array}
\usepackage{graphicx}

\def\be{\begin{equation}}
\def\ee{\end{equation}}
\def\bea{\begin{eqnarray}}
\def\eea{\end{eqnarray}}

\begin{document}

\title{Extrinsic and intrinsic curvatures in thermodynamic geometry}%

\author{Seyed Ali Hosseini Mansoori$^{1,2}$}
\email{shossein@bu.edu \& sa.hosseinimansoori@ph.iut.ac.ir}
\author{Behrouz Mirza$^{2}$}
\email{b.mirza@cc.iut.ac.ir}
\author{Elham Sharifian$^{2}$}
\email{e.sharifian@ph.iut.ac.ir}
\affiliation{$^1$Department of Physics, Boston University, 590 Commonwealth Ave., Boston, MA 02215, USA\\
$^{2}$Department of Physics,
Isfahan University of Technology, Isfahan 84156-83111, Iran}

\date{\today}%

\begin{abstract}
 We investigate the intrinsic and extrinsic curvature of a certain hypersurface in thermodynamic geometry of a physical system and show that they contain useful thermodynamic information.  For an anti-Reissner-Nordstr\"{o}m-(A)de Sitter black hole (Phantom), the extrinsic curvature of a constant $Q$ hypersurface has the same sign as the heat capacity around the phase transition points.
The intrinsic curvature of the hypersurface can also be divergent at the critical points but has no information about the sign of the heat capacity. Our study explains the consistent relationship holding between the thermodynamic geometry of the KN-AdS black holes and those of the RN ($J$-zero  hypersurface) and Kerr black holes ($Q$-zero hapersurface) ones \cite{ref1}. This approach can easily be generalized to an arbitrary thermodynamic system.

\end{abstract}


\maketitle

\section{Introduction}\label{a}
Bekenstein and Hawking showed that a black hole has a  behavior similar to a common thermodynamic system \cite{ref2, ref3}. They drew a parallel relationship between the four laws of thermodynamics and the physical properties of black holes by considering the surface gravity and the horizon area as the temperature and entropy, respectively \cite{ref4}. An interesting topic is to study phase transitions in black hole thermodynamics where the heat capacity diverges \cite{ref5, ref6}.  These divergence points of heat capacity are usually associated with a second order phase transition for some fixed black hole parameters \cite{ref24}.

Geometric concepts can also be used to study the  properties of an equilibrium space of thermodynamic systems. Riemannian geometry in the space of equilibrium states was introduced by Weinhold \cite{ref8} and Ruppeiner \cite{ref9,ref10} who defined metric elements as the Hessian matrix of the internal
energy and  entropy. These geometric structures are used to find the significance of the distance between equilibrium states. Consequently, various thermodynamic properties of the system can be derived from the properties of these metrics, especially critical behaviors, and stability of various types of black hole families \cite{ref11, ref12, ref13}. For the second order phase transitions, Ruppeiner curvature scalar (R) is expected to diverge at  critical points \cite{ref14,ref24,ref15,ref16, ref17}. Due to the success of this geometry to identify a phase transition,  several works \cite{ref18,ref19,ref20,ref21} have exploited it to  explain the black hole phase transitions.

The Ruppeiner geometry has also been analyzed for several black holes to find out the thermodynamic properties \cite{ref22, ref23}. As a result, the Ruppeiner curvature is flat for the BTZ and Reissner-Nordstr\"{o}m (RN) black holes, while curvature singularities occur for the Reissner-Nordstr\"{o}m anti de Sitter (RN-AdS) and Kerr black holes. Moreover, it has been argued in \cite{ref1} that all possible physical fluctuations could be considered for calculating curvature because
neglecting one parameter may lead to inadequate information about it. Therefore, the thermodynamic curvature of RN should be reproduced from the Kerr-Newmann anti-de Sitter (KN-AdS) black hole when the angular momentum $J\to 0$ and cosmological constant $\Lambda \to 0$. This approach  leads to a non-zero value for the Ruppeiner scalar, which is in contrast to the reports on RN in pervious works \cite{ref22, ref23}.

The present letter seeks to explain  this contrast by obtaining intrinsic and extrinsic curvatures of the related submanifolds. The induced metric (intrinsic curvature) and the extrinsic curvature of a constant $J$ hypersurface contain the necessary information about the properties of this hypersurface. The zero limit of an angular momentum for a KN-AdS black hole is equivalent to the two-dimensional constant J hypersurface embedded in a three-dimensional complete thermodynamic space. The curvature scalar of KN-AdS black hole on this hypersurface can be decomposed into an intrinsic curvature (Ruppeiner curvature of RN black hole), which is zero, and an extrinsic part that give the curvature singularities.

 We also prove that there is a one-to-one correspondence between divergence points of the heat capacities and those of the extrinsic curvature for thermodynamic descriptions where potentials are related to the mass (rather than  the entropy) by Legendre transformations. In spite of this correspondence, we can get other information about thermodynamics like stability and non-stability regions around phase transitions from singularities of extrinsic curvature and certain elements of the Ricci tensor. 

The organization of the letter is as follows. In Section \ref{S1} and \ref{S5}, we  analyze the nature of the phase transition through the diagrams of the Riemann tensor elements and extrinsic curvature. In Section \ref{S3}, we try to provide an answer
 to the question arising from the article \cite{ref1}, \textit{'' Ruppeiner geometry of RN black holes: flat or curved?"} using the concept of thermodynamic hypersurface in lower dimensions. 
 In Section \ref{S6}, we consider a Pauli paramagnetic gas and investigate a hypersurface in its thermodynamic geometry that corresponds to a zero magnetic field.
 Section
\ref{S4} contains a discussion of our results.


  \section{Thermodynamic extrinsic curvature } \label{S1}
We begin with a  review of our previous results on the correspondence between second order phase transitions and singularities of the thermodynamic geometry \cite{ref20, ref21}. We also introduce extrinsic curvature as a new concept of  the thermodynamic geometry. We will use this quantity in determining some information about stability and non-stability regions around phase transitions.
For charged black holes, a specific heat at a fixed electric charge is defined as follows:
\begin{equation}\label{h21}
{{C}_{Q}}=T{{\left( \frac{\partial S}{\partial T} \right)}_{Q}}=Tdet \left[\frac{\partial(S,Q)}{\partial{(T,Q)}} \right]=\frac{T{{\left\{ S,Q \right\}}_{S,Q}}}{{{\left\{ T,Q \right\}}_{S,Q}}}
\end{equation}
It is obvious that the phase transitions of ${{C}_{Q}}$ are the zeros of ${{\left\{ T,Q  \right\}}_{S,Q}}$  (Appx. \ref{A1} may be consulted  for a brief introduction to the bracket notation). Moreover, the Ruppeiner metric in the mass representation can be expressed as:
\begin{equation}\label{ee13}
{\textbf{g}}^{R}=\frac{H_{i,j}{M}}{T}
\end{equation}
where $H_{i,j}M=\left( {{{\partial }^{2}}{M}}/{\partial {{X}^{i}}\partial {{X}^{j}}} \right)$ is called the Hessian matrix and ${{X}^{i}}=(S,Q)$  are extensive parameters. Therefore, according to the first law  of thermodynamics, ${dM}={T dS+{\Phi} dQ}$, one could define the denominator of the scalar curvature
 $R(S,Q)$ by:
\begin{equation}\label{eeeee1}
g=det({\textbf{g}}^{R})=det \left[\frac{\partial(T,\Phi)}{T\partial{(S,Q)}} \right]=\frac{{{\left\{ T,\Phi  \right\}}_{S,Q}}}{{{T}^{2}}} =\frac{1}{T{{C}_{\Phi }}{{C}_{S}}}
\end{equation}
where, ${{C}_{S}}\equiv {{\left( {\partial Q}/{\partial \Phi } \right)}_{S}}$  and,
\begin{equation}\label{e1}
{{C}_{\Phi }}=T{{\left( \frac{\partial S}{\partial T} \right)}_{\Phi }}=\frac{T{{\left\{ S,\Phi  \right\}}_{S,Q}}}{{{\left\{ T,\Phi  \right\}}_{S,Q}}}
\end{equation}
 As a result, the scalar curvature $R(S,Q)$ is not able to explain the properties of the phase transitions of $C_{Q}$. From Eq. (\ref{eeeee1}), it is obvious that the phase transitions of ${{C}_{\Phi }}$ correspond precisely to the singularities of $R(S,Q)$. Now, one is able to prove an exact correspondence between singularities of this new metric $(\bar{R}(S, \Phi))$ and phase transitions of $C_Q$ \cite{ref20} by redefining the Ruppeiner metric as follows:
 \begin{equation} \label{ee12}
 \overline{\textbf{{g}}}=\frac{H_{i,j}{\overline{M}}}{T}
 \end{equation}
 where, $\overline{M}(S,\Phi )=M(S,Q)-\Phi Q$  is the enthalpy potential for $M(S,Q)$ and  ${{X}^{i}}=(S,\Phi  )$. From the first law, $
{d\overline{M}}(S,\Phi )=T dS-\Phi dQ$,
the denominator of $\overline{R}(S,\Phi )$ is obtained by:
\begin{equation} \label{h22}
\overline{g}=det({\overline{\textbf{g}}})=det \left[\frac{\partial(T,-Q)}{T\partial{(S,\Phi)}} \right]=-\frac{1}{{{T}^{2}}}{{\left\{ T,Q \right\}}_{S,\Phi }}=-\frac{C_S}{TC_Q}
\end{equation}
It is straightforward to show that the phase transitions of ${{C}_{Q}}$ are equal to the singularities of $\overline{R}(S,\Phi )$.
  We now examine a relationship between the divergences of the extrinsic curvature and the phase transition points. As already mentioned, the extrinsic curvature can be constructed by living on a certain hypersurface with a normal vector (See Appx. \ref{A2}). Since the heat capacity, $C_{Q}$, is defined at a constant electric charge, we should set on a constant $Q$ hypersurface. To do this, we change the coordinate from $(S,\Phi)$ to $(S,Q)$ by using the following Jacobian matrix.
 \begin{equation}\label{eq1}
 \textbf{J}\equiv \frac{\partial \left( S,\Phi  \right)}{\partial \left( S,Q \right)}
 \end{equation}
 The metric elements of  ${\overline{M}}(S,\Phi)$ in the new coordinate $(S,Q)$ can also be changed as follows:
 \begin{equation}\label{aa1}
 \overline{g}_{ij}'=J_{ik}^{T} \,\ \overline{g}_{kl} \,\ {{J}_{lj}}
 \end{equation}
 where, $J^{T}$ is the transpose of $J$. One can also rewrite  Eq. (\ref{ee12}) as a Jacobian matrix by:
 \begin{equation}\label{eq2}
 \overline{\textbf{g}}=\frac{\partial \left( T,-Q \right)}{T\partial \left( S,\Phi  \right)}
 \end{equation}
 Thus, the new metric takes the following form:
 \begin{eqnarray}\label{eq3}
\nonumber {{\bar{{\textbf{g}}'}}}&=&{{\left( \frac{\partial \left( S,\Phi  \right)}{\partial \left( S,Q \right)} \right)}^{T}}\left( \frac{\partial \left( T,-Q \right)}{T\partial \left( S,\Phi  \right)} \right)\left( \frac{\partial \left( S,\Phi  \right)}{\partial \left( S,Q \right)} \right)\\
 &=&{{\left( \frac{\partial \left( S,\Phi  \right)}{\partial \left( S,Q \right)} \right)}^{T}}\left( \frac{\partial \left( T,-Q \right)}{T\partial \left( S,Q \right)} \right)
 \end{eqnarray}
Furthermore by
 regarding, given the property of the determinant, i.e., $det(J^{T})=det(J)$, the determinant of the above relation can be calculated as follows:
 \begin{eqnarray}\label{eq4}
\nonumber \bar{{{g}}'}=\det \left( {{\left( \frac{\partial \left( S,\Phi  \right)}{\partial \left( S,Q \right)} \right)}^{T}}\left( \frac{\partial \left( T,-Q \right)}{\partial \left( S,Q \right)} \right) \right)=\\
 \det \left( \frac{\partial \left( S,\Phi  \right)}{\partial \left( S,Q \right)} \right)\det \left( \frac{\partial \left( T,-Q \right)}{\partial \left( S,Q \right)} \right)=-\frac{{{C}_{S}}^{-1}}{T{{C}_{Q}}}
\end{eqnarray}
 On the other hand, when we restrict ourselves to live on the constant $Q$ hypersurface with a normal vector ${{\overline{n}}_{Q}}=-1/\sqrt{\left| {{\overline{g}}'^{Q Q}} \right|}$, the extrinsic curvature will be given by:
 \begin{equation}
\overline{K}(S,Q )=\frac{1}{2\bar{g'}}\left[ \left( {{\overline{n}}^{\mu }}{{\partial }_{\mu }} \right)\bar{g'}\right]+\left( {{\partial }_{\mu }}{{\overline{n}}^{\mu }} \right)
 \end{equation}
 where, ${{\overline{n}}^{\mu }}=\left( {{\overline{n}}^{S}},{{\overline{n}}^{Q }} \right)=\left( {{\overline{g'}}^{SQ }},{{\overline{g'}}^{QQ}} \right){{\overline{n}}_{Q }}$ (See Appx. \ref{A2}). From Eq. (\ref{eq3}), the metric tensor in the new coordinate can be calculated as follows: 
 \begin{equation}
 {{\bar{{\textbf{g}}'}}}=diag(C_{Q}^{-1},-\frac{C_{S}^{-1}}{T})
 \end{equation}
Therefore, on the constant $Q$ hypersurface, we have: 
 \begin{equation}
  {{\overline{n}}^{\mu }}=\left( 0,{{\overline{n}}^{Q}} \right)=\left( 0,\sqrt{\left| T{{C}_{S}} \right|} \right)
 \end{equation}
 It is easy to show that the above vector is a normalized vector ($\bar{n}_{\mu} \bar{n}^{\mu}=-1$). Therefore, the extrinsic curvature can be rewritten as follows:
 \begin{equation}
 \bar{K}={{\left. \frac{T{{C}_{Q}}}{{{C}_{S}^{-1}}}\left[ \sqrt{\left| T{{C}_{S}} \right|}{{\partial }_{Q}}\left( \frac{{{C}_{S}^{-1}}}{T{{C}_{Q}}} \right) \right]+{{\partial }_{Q}}\sqrt{\left| T{{C}_{S}} \right|} \right|}_{Q=cte}}
 \end{equation}
 Indeed this relation tells us that the singularities of this curvature occur exactly at phase transitions of the $C_{Q}$. It should be noted that in this case when the extrinsic curvature diverges,  the metric components are differentiable. However, the metric elements are non-differentiable for extremal black holes in the Ruppeiner geometry.  Our study indicates that in thermodynamic geometry divergences of the extrinsic curvature does not always implies non-differentiavble metric elements.
 In table I, we compare the heat capacity and the extrinsic curvature function for $Kerr$, $RN$, $BTZ$, and Einstein-Maxwell-Gauss-Bonnet ($EMGB$) \cite{Ref6} black holes. In all cases, the roots of the extrinsic curvature denominator show phase transition points. The extrinsic curvature also changes its sign 
at the phase transition points which is exactly a similar behavior to heat capacity. 
 Generally, for thermodynamic systems with $(n+1)$ variables, one could consider the following metric,
 \begin{equation}
 \overline{\textbf{g}}=\frac{H_{i,j}\overline{M}}{T}=\frac{\partial \left( T,-Q_{1} ,-Q_{2},...,Q_{n} \right)}{T\partial \left( S,\Phi_{1} ,\Phi_{2},...,\Phi_{n}\right)}
 \end{equation}
 where $\overline{M}=M-\sum_{i}^{n}\Phi_{i}Q_{i}$ and $X^{i}=\left(S,Q_{1},...,Q_{n}\right)$. Furthermore, $\left(T,\Phi_{1},...,\Phi_{n}\right)$ are called extensive parameters.
 Then utilizing the Jacobian matrix as follows:
 \begin{equation}
 \textbf{J}\equiv \frac{\partial \left( S,\Phi_{1},\Phi_{2},...,\Phi_{n}  \right)}{\partial \left( S,Q_{1},Q_{2},...,Q_{n} \right)}
 \end{equation}
 in a similar way, the metric tensor can be represented by below block-diagonal matrix.
 \begin{equation}
 \bar{{{\textbf{g}}}'}=diag(C_{Q_{1},Q_{2},...,Q_{n}}^{-1},-\textbf{G} )
 \end{equation}
 where $G$ is a square matrix of order $n$ defined by the following relation,
 \begin{equation}
 \textbf{G}=\frac{H_{i,j}{M}}{T} \,\ ; \,\ X^{i}=(Q_{1},Q_{2},...,Q_{n})
 \end{equation}
 Thus the metric determinant in the new coordinates can be written as:
 \begin{equation}
 \bar{{{g}}'}= \frac{\left[C_{S,Q_{1},Q_{2},...,Q_{n}}C_{S,\Phi_{1},Q_{2},...,Q_{n}}...C_{S,\Phi_{1},\Phi_{2},...,\Phi_{n}}\right]^{-1}}{{(-T)}^{n}C_{Q_{1},Q_{2},...,Q_{n}}}
 \end{equation}
where above functions were defined in \cite{ref20}. Now, a constant $Q_{i}$ hypersurface has the following unit normal vector,
\begin{equation}
\bar{n}_{\mu}=(0,0,...,-{(T C_{S,\Phi_{1},...,\Phi_{i-1},\Phi_{i+1},...,\Phi_{n}})^\frac{-1}{2}},..,0)
\end{equation}
 where the non-zero term places in $i^{th}$ column. The extrinsic curvature, $\bar{K}(S,Q_{1},Q_{2},...,Q_{n})$, can be calculated using Eq (\ref{ee8}). It is interesting that the extrinsic curvature has the same behavior as the specific heat, 
 $C_{Q_{1},Q_{2},...,Q_{n}}$. 
 
\begin{figure}[tbp]
 \includegraphics[scale=.4]{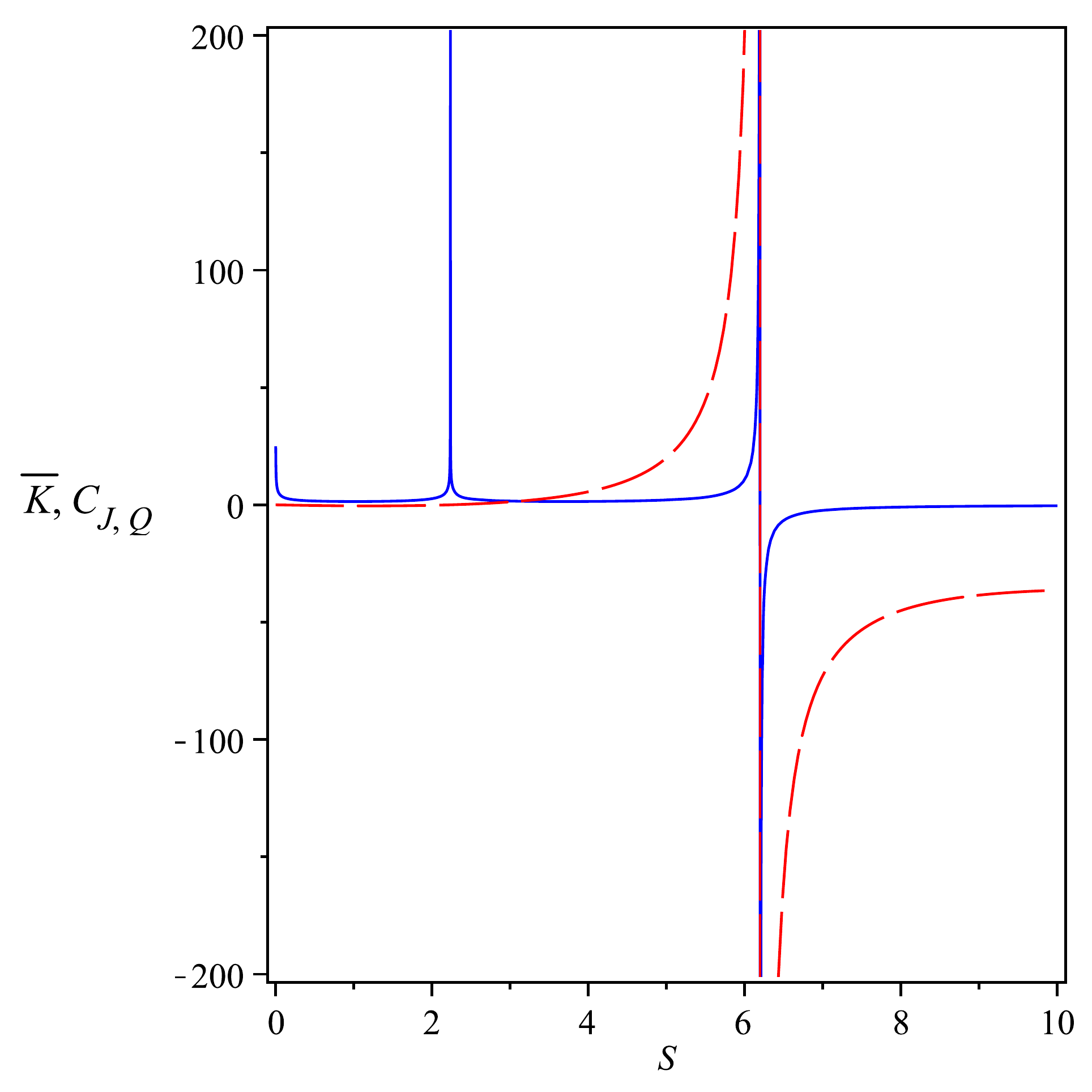}
 \caption{Graph of the extrinsic curvature $\bar{K}$ (solid blue curve) and the heat capacity $C_{J,Q}$ (dashed red curve) as a  function of  entropy, $S$, for
 an electric charge $Q=1$ and an angular momentum $J=1$.\label{fig4}   }
 \end{figure}

 For a Kerr-Newman ($KN$) black hole with the following mass,
 \begin{equation}
 M=\frac{\sqrt{S(4{{J}^{2}}+{{S}^{2}}+2{{Q}^{2}}S+{{Q}^{4}})}}{2S}
 \end{equation}
 metric elements are defined as follows:
\begin{equation}
\overline{\textbf{g}}=\frac{H_{i,j}\overline{M}}{T}=\frac{\partial \left( T,-Q ,-J \right)}{T\partial \left( S,\Phi ,\Omega\right)}
\end{equation}
where $T$ is the Hawking temperature, $\Omega$ is the angular velocity, and  $\Phi$ is the potential deference \cite{ref20,ref21}. 
 When somebody restricts himself to live on the constant $J$ hypersurface which has the orthogonal normal vector, ${{\overline{n}}_{J}}=-1/\sqrt{\left| {TC_{S,\Phi}} \right|}$, the extrinsic curvature diverges at the phase transition point and exhibit a similar sign behavior around the  transition points. In Figure \ref{fig4}, the graph of the extrinsic curvature, $\overline{K}$, and the heat capacity, $C_{J,Q}$, for the Kerr-Newman black hole ($KN$) shows  an exact correspondence between singularities and phase transitions (Note that the first divergence point is related to $T=0$.). It is surprising that  the same result obtains by considering a constant $Q$ hypersurface with unit normal vector, ${{\overline{n}}_{Q}}=-1/\sqrt{\left| {TC_{S,\Omega}} \right|}$.
Moreover, it will be easy to show that the signs of such Ricci tensor elements as $\overline{R}_{SS}$, $\overline{R}_{SQ}$ and $\overline{R}_{SJ}$ are similar to that of the $C_{J, Q}$ around the transition points. 
  
 \begin{table*}\label{aa}
\begin{tabular*}{\textwidth}{@{\extracolsep{\fill}}lrrl@{}}
\hline
\\
$kerr$ & $RN$ \\
\\
\hline
\\
$M(S,J)=\frac{\sqrt {{\frac {S}{\pi }}+4\,{\frac {{J}^{2}\pi }{S}}}}{2} $ & $M(S,Q)=\frac{\sqrt {S\pi } \left( {\pi }^{-1}+{\frac {{Q}^{2}}{S}} \right)}{2}$  \\
\\
$T(S,J)=-{\frac {4\,{J}^{2}{\pi }^{2}-{S}^{2}}{4\sqrt {\pi }{S}^{2}}{
\frac {1}{\sqrt {{\frac {4\,{J}^{2}{\pi }^{2}+{S}^{2}}{S}}}}}}$ & $T(S,Q)=-{\frac {\sqrt {S\pi }{Q}^{2}}{2{S}^{2}}}$\\
\\
$\overline{K}(S,J)={\frac {2J\pi \, \left( 48\,{J}^{4}{\pi }^{4}+8\,{J}^{2}{\pi }^{2}{S
}^{2}+3\,{S}^{4} \right) \sqrt {2}\sqrt {S}}{ \left( 48\,{J}^{4}{\pi }
^{4}+24\,{J}^{2}{\pi }^{2}{S}^{2}-{S}^{4} \right) \sqrt {16\,{J}^{4}{
\pi }^{4}-{S}^{4}}}}$ & $\overline{K}(S,Q)=0$ \\
\\
$C_{J}(S,J)=-{\frac {2S \left( 16\,{J}^{4}{\pi }^{4}-{S}^{4} \right) }{48\,{J}^{
4}{\pi }^{4}+24\,{J}^{2}{\pi }^{2}{S}^{2}-{S}^{4}}}$ & $C_{Q}(S,Q)=-S/2$\\
\\
\hline
\\
$BTZ$ & $EMGB$ \\
\\
\hline
\\
$M(S,J)={\frac {{S}^{2}}{16{\pi }^{2}{l}^{2}}}+4\,{\frac {{\pi }^{2}{J}^{2
}}{{S}^{2}}}$ & $M(S,Q)=\pi \alpha +{\frac {\pi \,{Q}^{2}}{6\sqrt [3]{{S}^{2}}}}+{\pi }^{2
}\sqrt [3]{{S}^{2}}-\frac{\pi \,\Lambda\,\sqrt [3]{{S}^{4}}}{12}$\\
\\
$T(S,J)={\frac {S}{8{\pi }^{2}{l}^{2}}}-8\,{\frac {{\pi }^{2}{J}^{2}}{{S}^{3}}}$ & $T(S,Q)=-{\frac {\pi \, \left( \Lambda\,{S}^{4}\sqrt [3]{{S}^{2}}-6\,\pi
\, \left( {S}^{2} \right) ^{2/3} \left( {S}^{4} \right) ^{2/3}+{Q}^{2}
 \left( {S}^{4} \right) ^{2/3} \right) }{9S\sqrt [3]{{S}^{2}} \left( {S
}^{4} \right) ^{2/3}}}$\\
\\
$\overline{K}(S,J)={\frac {32{S}^{7/2}J{\pi }^{2}l}{ \left( 192\,{J}^{2}{\pi }^{4}{l}^{
2}+{S}^{4} \right) \sqrt {64\,{J}^{2}{\pi }^{4}{l}^{2}-{S}^{4}}}}$ & $\overline{K}(S,Q)={\frac {2Q{S}^{3/2} \left( \Lambda\,{S}^{2/3}-4\,\pi  \right) \sqrt
{3}}{ \left( \Lambda\,{S}^{8/3}-5\,{Q}^{2}{S}^{2/3}+6\,\pi \,{S}^{2}
 \right) \sqrt {\Lambda\,{S}^{2}-6\,\pi \,{S}^{4/3}+{Q}^{2}}}}$ \\
 \\
$C_{J}(S,J)=-{\frac { \left( 64\,{J}^{2}{\pi }^{4}{l}^{2}-{S}^{4} \right) S}{192\,
{J}^{2}{\pi }^{4}{l}^{2}+{S}^{4}}}$ & $C_{Q}(S,Q)={\frac {3S \left( \Lambda\,{S}^{8/3}+{Q}^{2}{S}^{2/3}-6\,\pi \,{S}^{
2} \right) }{\Lambda\,{S}^{8/3}-5\,{Q}^{2}{S}^{2/3}+6\,\pi \,{S}^{2}}}$\\
\\
\hline
\end{tabular*}
\caption{  Thermodynamic variables and extrinsic curvature functions for $Kerr$, $RN$, $BTZ$ and $EMGB$ black holes.}
\end{table*}

\section{Extrinsic curvature of the phantom RN-AdS black hole}
\label{S5}
\begin{figure}[tbp]
 \includegraphics[scale=.4]{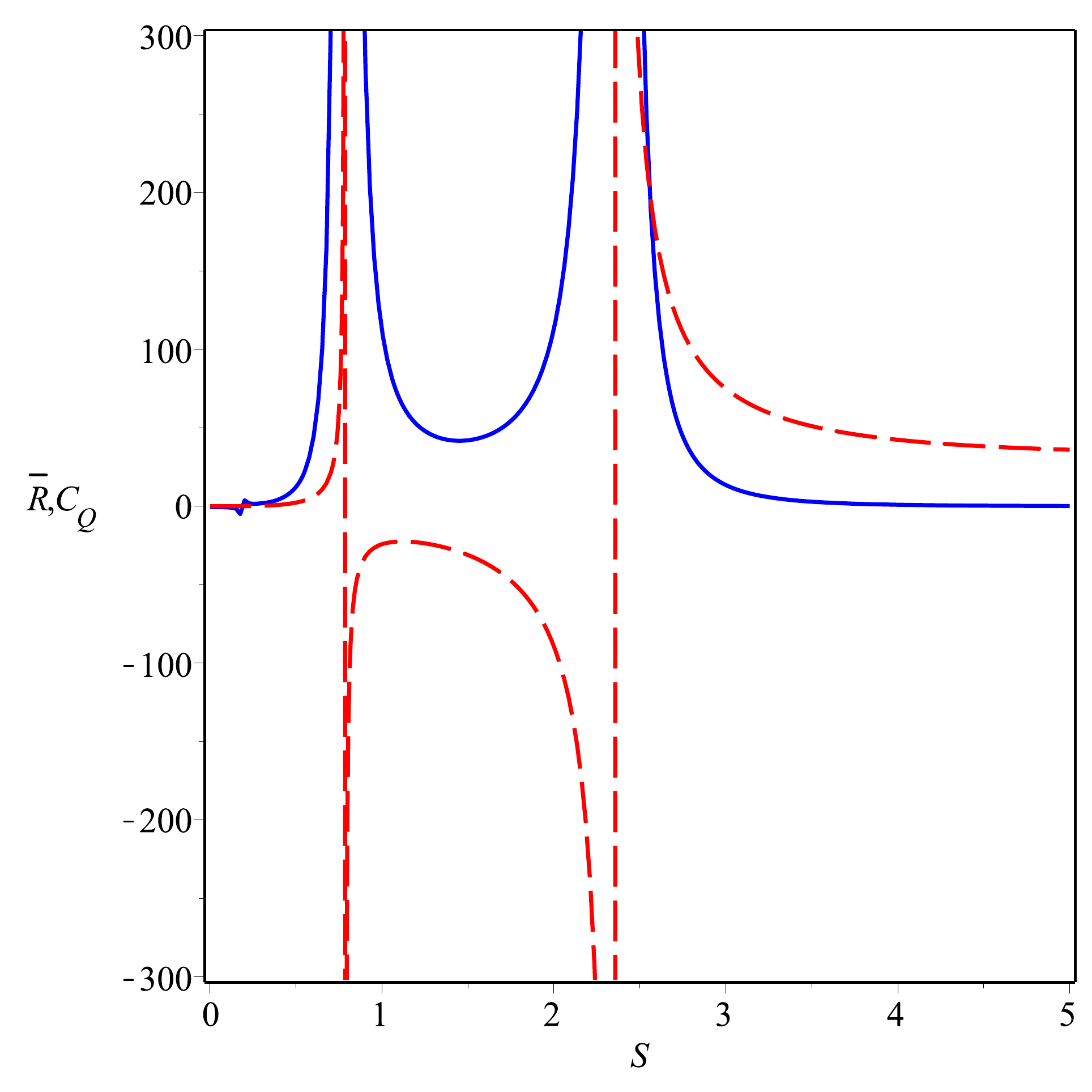}
 \caption{Graph of the scalar of curvature $\overline{R}$ (solid blue curve) and the heat capacity $C_{Q}$ (dashed red curve) as a  function of  entropy, $S$,  in the {\bf RN-AdS} case, for
 an electric charge $Q=0.25$ and a cosmological constant $\Lambda$=- 1.\label{fig1}  }

 \includegraphics[scale=.4]{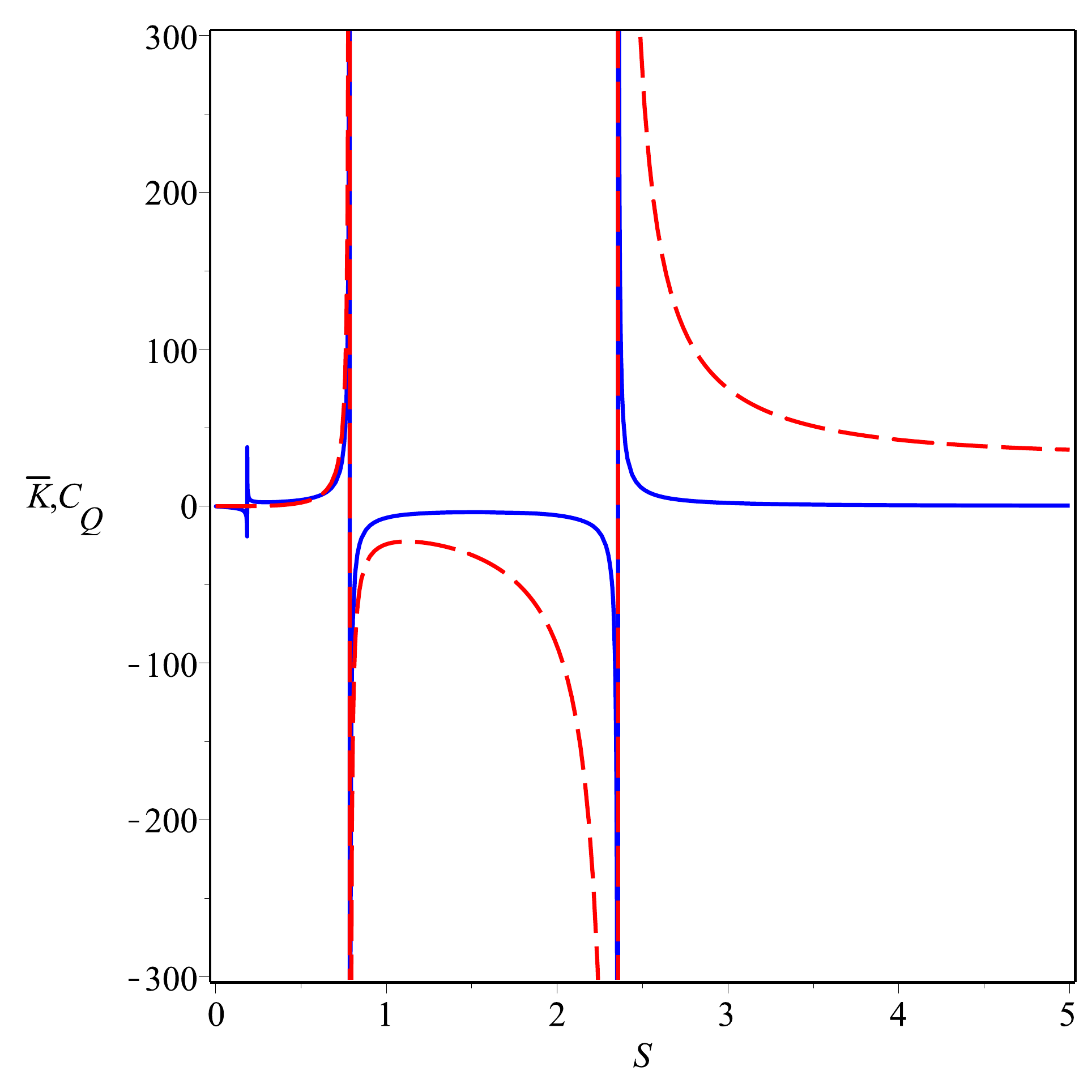}
 \caption{Graph of the scalar of curvature $\overline{K}$ (solid blue curve) and the heat capacity $C_{Q}$ (dashed red curve) as a  function of  entropy, $S$,  in the {\bf RN-AdS} case, for
 an electric charge $Q=0.25$ and a cosmological constant $\Lambda$ =- 1.\label{fig2}  }
 \end{figure}

\begin{figure}[tbp]
 \includegraphics[scale=.4]{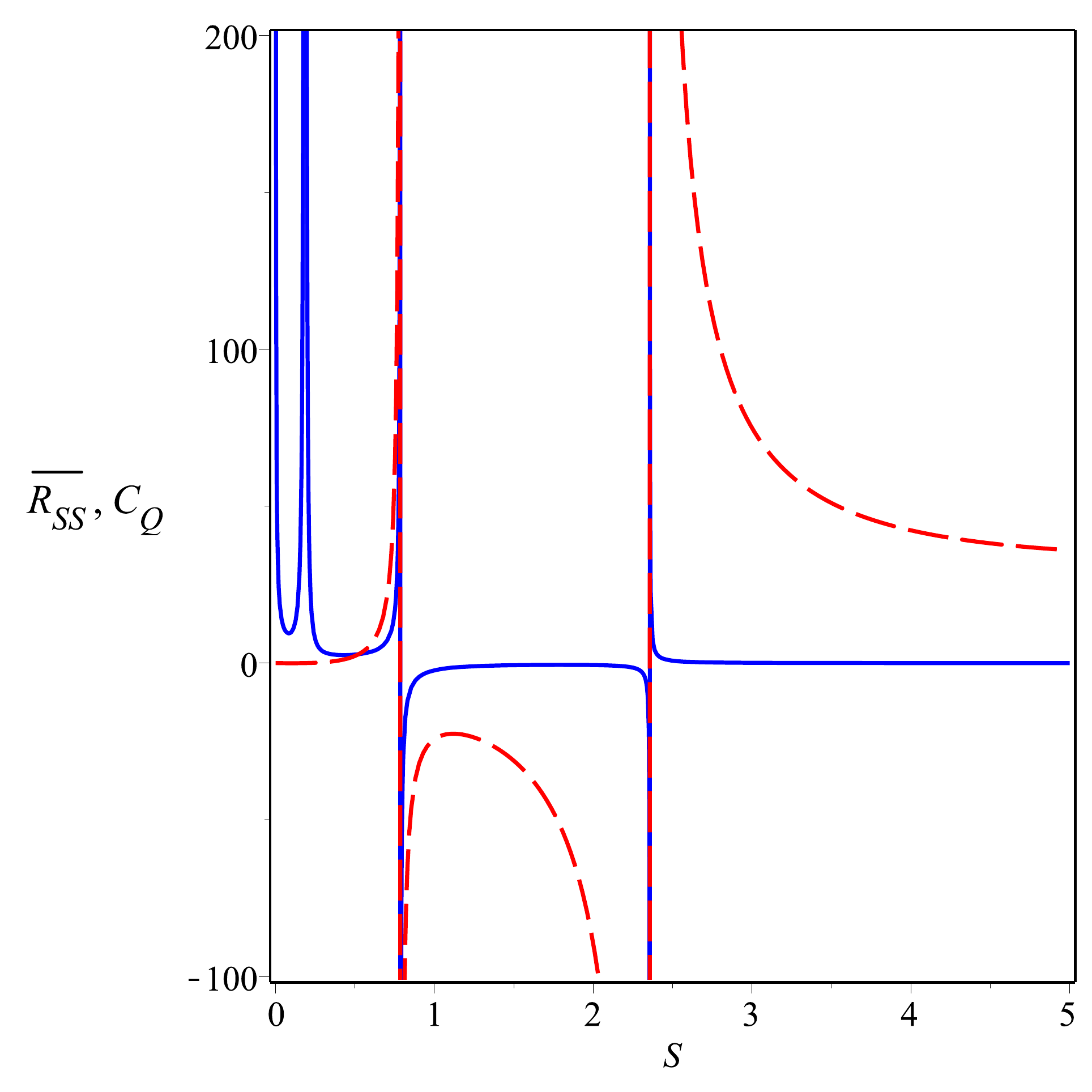}
 \caption{Graph of the scalar of curvature $\overline{R}_{SS}$ (solid blue curve) and the heat capacity $C_{Q}$ (dashed red curve) as a  function of  entropy, $S$,  in the {\bf RN-AdS} case, for
 an electric charge $Q=0.25$ and a cosmological constant $\Lambda$=- 1.\label{fig3}   }
 \end{figure}

The action for the Einstein-Hilbert theory with phantom Maxwell field reads:
\begin{equation}
S= \int \sqrt{-g} \left (R+2\Lambda + 2\eta F_{\mu\nu}F^{\mu\nu}\right) d^4 x\ ,
\end{equation}
  where, $\Lambda$ is the cosmological constant, and $\eta= \pm 1$. RN-AdS black hole corresponds to $\eta=1$, while phantom couplings of the Maxwell field ( Phantom RN-AdS black hole) are obtained for
$\eta=-1$. The metrics of these solutions,
derived in \cite{ref25}, take the below form.
\begin{equation}
ds^2 = f(r) dt^2 -\frac{1}{f(r)}dr^2 - r^2 (d\theta^2 + \sin^2\theta\, d\phi^2)\ ,
\label{pbhmet}
\end{equation}
where $f(r)$ is given by:
\begin{equation}
f(r) = 1 - \frac{2M}{r} -\frac{\Lambda}{3}r^2 + \eta\frac{Q^2}{r^2}
\label{pbhsol}
\end{equation}
The event horizon, $r_{+}$, of this solution can be determined by calculating the roots for the equation $f(r_{+}) = 0$. 
The mass of this black hole is expressed as a function of the thermodynamic variables.
 \begin{equation} \label{ss2}
  M=\frac{1}{2}{{(S/\pi )}^{3/2}}(\frac{\pi }{S}-\frac{\Lambda }{3}+\frac{\eta {{\pi }^{2}}{{Q}^{2}}}{{{S}^{2}}})\
 \end{equation}
 where $S=\pi r_{+}^{2}$ is the Bekenstein-Hawking entropy. According to the first law of thermodynamics, one can calculate the Hawking temperature,  $T$,  the electric potential,  $\Phi$,  and the specific heat capacity, $C_{Q}$, as follows:
 \begin{equation} \label{mm2}
 T={(\frac{\partial M}{\partial S})}_{Q}=\frac{-\pi S+\Lambda {{S}^{2}}+\eta {{\pi }^{2}}{{Q}^{2}}}{-4{{(\pi S)}^{3/2}}}\
 \end{equation}
 \begin{equation}
 \Phi ={(\frac{\partial M}{\partial Q})}_{S}=\frac{{{(S/\pi )}^{3/2}}\eta {{\pi }^{2}}Q}{{{S}^{2}}}\
 \end{equation}
 \begin{equation}  \label{ee6}
 {{C}_{Q}}=T{{(\frac{\partial S}{\partial T})}_{Q}}=\frac{-2S(-\pi S+\Lambda {{S}^{2}}+\eta {{\pi }^{2}}{{Q}^{2}})}{(-\pi S-\Lambda {{S}^{2}}+3\eta {{\pi }^{2}}{{Q}^{2}})}
\end{equation}
 Using Eq. (\ref{ee12}), it is easy to obtain the
 metric elements of the enthalpy potential in the coordinate $(S, \Phi)$.  Then by applying Eqs. (\ref{eq1}) and (\ref{aa1}), the scalar curvature,  $\overline{R}$, takes the following form:
 \begin{equation} \label{ee5}
 \bar{R}=\frac{F(S,Q)}{(-S\pi  + \Lambda {{S}^{2}}+{{\pi }^{2}}\eta {{Q}^{2}}){{(-S\pi  - \Lambda {{S}^{2}}+3{{\pi }^{2}}\eta {{Q}^{2}})}^{2}}}
 \end{equation}
 where,
 \begin{eqnarray}
& \nonumber F=-{\eta}^{2}{Q}^{4}{\pi }^{5}+ \left( 2\,\eta\,{Q}^{2}+10\,\Lambda\,{
\eta}^{2}{Q}^{4} \right) S{\pi }^{4}+2\,{\Lambda}^{2}{S}^{4}\pi\\
&+{S}^{2} \left( -9\,\eta\,{Q}^{2}
\Lambda-1 \right) {\pi }^{3}+ \left( 3\,\Lambda-6\,{\Lambda}^{2}\eta\,
{Q}^{2} \right) {S}^{3}{\pi }^{2}
 \end{eqnarray}
 The first part of the denominator is zero only at $T=0$ and the roots of the second part of the curvature denominator gives us the phase transition points. Therefore, the curvature diverges exactly at these points where  heat capacity diverges with no other additional roots. For the RN-AdS black hole, the scalar curvature (\ref{ee5}) and the specific heat (\ref{ee6}) are depicted in Figure \ref{fig1} as a function of entropy and for a fixed value of  electric charge $Q = 0.25$.
On the other hand,
 the extrinsic curvature opens an interesting and impressive avenue to the  investigation of how phase transitions behave. In this case, we need to sit on a hypersurface with the normal vector ${{\overline{n}}_{Q}}=-1/\sqrt{\left| {{{\bar{g}}}^{QQ}} \right|}$
 in which the extrinsic curvature associated with this hypersurface is given by:
 \begin{equation}
 \bar{K}=-{\frac { \left( \pi -2\,\Lambda\,S \right) \sqrt {\left|\eta \right|}\pi \,Q\sqrt {
S}}{ \left( -\pi \,S-\Lambda\,{S}^{2}+3\,\eta\,{\pi }^{2}{Q}^{2}
 \right) \sqrt {\left|-\pi \,S+\Lambda\,{S}^{2}+\eta\,{\pi }^{2}{Q}^{2}\right|}}}
\end{equation}
The first term of the denominator shows the phase transition points, while the second is only zero at $T=0$. Interestingly, we see in Figure \ref{fig2} that the extrinsic curvature has the same sign as heat capacity does, while in Figure \ref{fig1}, the scalar curvature does not have the same sign as the heat capacity around the phase transition points. In other words, extrinsic curvature reveals  more information such as the stability/non-stability of heat capacities  than the Ricci scalar does. Figure \ref{fig3} indicates that the special Ricci tensor elements $\overline{R}_{SS}$ also exhibit a similar behavior around phase transition points such as heat capacity. Therefore, the extrinsic curvature $\overline{K}(S,Q)$ and the $\overline{R}_{SS}$ component of the Ricci tensor describe the phase transitions and the sign of the heat capacity, $C_{Q}$. Our work is a new method to identify stable regimes 
in the parameter space of black holes by studying the extrinsic curvature in their thermodynamic geometry.  

\section{Ruppeiner curvature of RN black hole as an intrinsic curvature on a constant $J$ hypersurface}\label{S3}
In article \cite{ref1}, the authors proposed a new measure of microscopic interactions and its effects on Ruppeiner curvature by considering a complete phase space of extensive variables. They obtained a new non-zero Ruppeiner curvature for RN black holes by setting $l\to \infty,\left({l=\frac{-3}{\Lambda}}\right)$, where $J\to 0$ limits in the scalar curvature for the Kerr-Newman-AdS (KN-AdS) black hole as follows:
\begin{equation}\label{eqq1}
\lim_{l \rightarrow \infty ; J \rightarrow 0 }{{R}_{KN-AdS}}=\frac{{{S}^{2}}+{{Q}^{2}}S+2{{Q}^{4}}}{{{\left( S+{{Q}^{2}} \right)}^{2}}\left( {{Q}^{2}}-S \right)}
\end{equation}
This non-vanishing scalar curvature is the result of another dimension specified by $J$ which fluctuates even if we set it to zero.
It is surprising that this result is in contrast to a direct calculation on the Ruppeiner metric of the $RN$ which is zero \cite{ref23}. One of our objectives in this work is to explain the difference between these two results by applying the basic concepts of the extrinsic/ intrinsic geometry for a particular hypersurface (See Appx. \ref{A2}). In this framework, the Ruppeiner curvature of the KN-AdS black hole can be broken down  into a purely intrinsic part, which yields a zero Ruppeiner curvature of the $RN$ black hole, and an extrinsic part, which measures the bending of the constant $J$ hypersurface; that is:
\begin{equation}\label{eqq5}
{}^{(3)}{{R}_{KN-AdS}}={}^{(2)}R_{in}-({{K}^{2}}-{{K}_{a b}}{{K}^{a b}})-2{{(n_{;\beta }^{\alpha }{{n}^{\beta }}-{{n}^{\alpha }}n_{;\beta }^{\beta })}_{;\alpha }}
\end{equation}
where, ${}^{(2)}R_{in}=R_{RN}=0$. Here  the extrinsic part is expected to be  exactly the same as Eq. (\ref{eqq1}).
Now, let us investigate  the accuracy of our claim about the KN-AdS black hole in the limits $l \to \infty$ and $J \to 0$.

The mass relation of the KN-AdS black hole \cite{ref26} as a function of thermodynamic variables can be written as:
\begin{eqnarray}
 M=[\frac{S}{4\pi }+\frac{\pi \left( 4{{J}^{2}}+{{Q}^{2}} \right)}{4S}+\frac{{{Q}^{2}}}{2}+\frac{{{J}^{2}}}{{{l}^{2}}}\\
 \nonumber +\frac{S}{\pi {{l}^{2}}}\left( {{Q}^{2}} +\frac{S}{\pi }+\frac{{{S}^{2}}}{2{{\pi }^{2}}{{l}^{2}}} \right)]^{\frac{1}{2}}
\end{eqnarray}
The Hawking temperature $T$ is also defined by:
\begin{eqnarray}
T={{\left( \frac{\partial M}{\partial S} \right)}_{Q,J}}=\frac{{{S}^{2}}{{\pi }^{2}}{{l}^{4}}-4{{\pi }^{4}}{{l}^{4}}{{J}^{2}}-{{\pi }^{4}}{{l}^{4}}{{Q}^{2}}}{8{{\pi }^{3}}M{{S}^{2}}{{l}^{4}}}\\
 \nonumber +\frac{2{{S}^{2}}{{Q}^{2}}{{\pi }^{2}}{{l}^{2}}+4{{S}^{3}}\pi {{l}^{2}}+3{{S}^{4}}}{8{{\pi }^{3}}M{{S}^{2}}{{l}^{4}}}
\end{eqnarray}
It should be noted that, within the hypersurface framework, setting $J$ to zero is tantamount to living  on the constant $J$ hypersurface ($J$- zero hapersurface) which has the following normal vector,
\begin{equation}\label{eqq4}
 {{n}_{J}}=-1/\sqrt{\left| {{{{g}}}^{JJ}} \right|} \,\,\ ; \,\,\ {{g}_{JJ}}=\frac{1}{T}\left( \frac{{{\partial }^{2}}M}{\partial {{J}^{2}}} \right).
\end{equation}
 Therefore, by making use of Eqs. (\ref{ee8}) and (\ref{eee1}), we have:
 \begin{equation}
 K^2=K_{\alpha \beta} K^{\alpha \beta}=0
 \end{equation}
On the other hand, the metric elements induced on the constant $J$ hypersurface can be given by:
\begin{eqnarray}
g_{SS}=\lim_{l \rightarrow \infty ; J \rightarrow 0 }{\frac{1}{T}\left( \frac{{{\partial }^{2}}M}{\partial {{S}^{2}}} \right)}=\frac{3{{Q}^{2}}-S}{2S\left( S-{{Q}^{2}} \right)}\\
g_{SQ}=g_{QS}=\lim_{l \rightarrow \infty ; J \rightarrow 0 }{\frac{1}{T}\left( \frac{{{\partial }^{2}}M}{\partial S\partial Q} \right)}=\frac{-2Q}{S-{{Q}^{2}}}\\
g_{QQ}=\lim_{l \rightarrow \infty ; J \rightarrow 0 }{\frac{1}{T}\left( \frac{{{\partial }^{2}}M}{\partial {{Q}^{2}}} \right)}=\frac{4S}{S-{{Q}^{2}}}
\end{eqnarray}
The above elements are the same as those of the Ruppeiner metric for the $RN$ black hole \cite{ref23}. Therefore, the intrinsic curvature ${}^{(2)}R_{in}$ of the constant $J$ hypersurface equals the Ruppeiner curvature of the $RN$ black hole; i.e.,
\begin{equation}
{}^{(2)}R_{in}=R_{RN}=0
\end{equation}
 Finally, based on Eq. (\ref{eqq4}), the last statement of Eq. (\ref{eqq5}) is given by:
\begin{equation}
\lim_{l \rightarrow \infty   ; J \rightarrow 0 }{{{(n_{;\beta}^{\alpha}{{n}^{\beta}}-{{n}^{\alpha}}n_{;\beta}^{\beta})}_{;\alpha}}}=-\frac{\frac{{{S}^{2}}}{2}+\frac{{{Q}^{2}}S}{2}+{{Q}^{4}}}{{{\left( S+{{Q}^{2}} \right)}^{2}}\left( {{Q}^{2}}-S \right)}
\end{equation}
We can, therefore,  conclude that the non-vanishing Ruppeiner curvature in the limits $J \to 0$ and $l \to \infty$ is extracted from the curvature of the KN-AdS black hole when one lives on the constant $J$ haypersurface. In addition, the Ruppeiner curvature of the RN black hole can be interpreted as the intrinsic curvature produced by  the induced metric on the two dimensional hypersurface ($J$-zero hypersurface).

 We may also check Eq. (\ref{ee11}) for the Kerr black hole at $Q \to 0$ and $l \to \infty$ limits of the KN-AdS black hole to show that the intrinsic part of Eq. (\ref{ee11}) is the Ruppeiner curvature of the Kerr black hole. Using the definition of the Ruppeiner metric for the complete phase space of the parameters (KN-AdS
black hole), and assuming  the limits $Q\to 0, \,\ l\to \infty $, we obtain the following equation for the Ricci scalar for a non- charged KN-AdS black hole.
\begin{equation}
\lim_{l \rightarrow \infty ; Q \rightarrow 0 }{{}^{(3)}{R}_{KN-AdS}}=\frac{\left( 36{{J}^{2}}+{{S}^{2}} \right)S}{16{{J}^{4}}-{{S}^{4}}}
\end{equation}
Also, the induced metric for the $Q$-zero hypersurface is calculated by:
\begin{widetext}
\begin{eqnarray}
  g_{SS}=\lim_{l \rightarrow \infty ; Q \rightarrow 0 }{\frac{1}{T}\left( \frac{{{\partial }^{2}}M}{\partial {{S}^{2}}} \right)}={\frac {-24{J}^{2}{S}^{2}-48\,{J}^{4}+{S}^{4}}{2S \left( 4\,{J}^
{2}+{S}^{2} \right)  \left( 4\,{J}^{2}-{S}^{2} \right) }}\\
  g_{SJ}=g_{JS}=\lim_{l \rightarrow \infty ; Q \rightarrow 0 }{\frac{1}{T}\left( \frac{{{\partial }^{2}}M}{\partial S\partial J} \right)}={\frac {4J \left( 4\,{J}^{2}+3\,{S}^{2} \right) }{ \left( 4\,{J}^{2}
+{S}^{2} \right)  \left( 4\,{J}^{2}-{S}^{2} \right) }}\\
  g_{JJ}=\lim_{l \rightarrow \infty ; Q \rightarrow 0 }{\frac{1}{T}\left( \frac{{{\partial }^{2}}M}{\partial {{J}^{2}}} \right)}={\frac {-8{S}^{3}}{ \left( 4\,{J}^{2}+{S}^{2} \right)  \left( 4\,{J}
^{2}-{S}^{2} \right) }}
\end{eqnarray}
\end{widetext}
One can obtain the following relation for the intrinsic curvature.
\begin{equation}
R_{in}=R_{kerr}={\frac { \left( 12\,{J}^{2}+{S}^{2} \right) S}{16\,{J}^{4}-{S}^{4}}}
\end{equation}
 Moreover,  for the $Q$-zero hypersurface with the normal vector,
\begin{equation}
 {{n}_{Q}}=-1/\sqrt{\left| {{{{g}}}^{QQ}} \right|} \,\,\ ; \,\,\ {{g}_{QQ}}=\frac{1}{T}\left( \frac{{{\partial }^{2}}M}{\partial {{Q}^{2}}} \right)
\end{equation}
the extrinsic curvature is zero and  $K^2=K_{\alpha \beta} K^{\alpha \beta}=0$.
 Now, we can successfully examine the validity of the following equation;
\begin{equation}
\lim_{l \rightarrow \infty ; Q \rightarrow 0 }{{R}_{KN-AdS}}={{}^{(2)}{R}_{Kerr}}  -   2\lim_{l \rightarrow \infty ; Q \rightarrow 0 }{{{(n_{;\beta}^{\alpha}{{n}^{\beta}}-{{n}^{\alpha}}n_{;\beta}^{\beta})}_{;\alpha}}}
\end{equation}
where,
\begin{equation}\label{eeee1}
\lim_{l \rightarrow \infty   ; Q \rightarrow 0 }{{{(n_{;\beta}^{\alpha}{{n}^{\beta}}-{{n}^{\alpha}}n_{;\beta}^{\beta})}_{;\alpha}}}=-12\,{\frac {S{J}^{2}}{16\,{J}^{4}-{S}^{4}}}
\end{equation}
In summary, the Ruppeiner curvature of the Kerr black hole is similar to the intrinsic curvature  produced by  the induced metric  on the two dimensional hypersurface ( i.e., constant $Q$ hypersurface). Our study indicates that in thermodynamic geometry,  properties of the intrinsic and the extrinsic curvatures are important to obtain a complete geometric representation of  
thermodynamics in physical systems.
The intrinsic curvature also help us to identify 
attractive, repulsive and non-interacting statistical interaction between the constituent parts of a thermodynamic system as we discuss in the next Section.
\section{Hypersurfaces and their intrinsic curvature  in thermodynamic geometry of the Pauli paramagnetic gas
}\label{S6}
  
Thermodynamic curvature may explain the statistical interaction between particles in a thermodynamic system \cite{Ref1,Ref2,Ref3}. The thermodynamic curvature is positive 
for attractive interaction between particles,
and negative for a repulsive interaction \cite{Ref3}.
In case that particles have not any interaction with each other, the thermodynamic geometry is flat \cite{Ref1}.
Now we consider thermodynamic geometry of a Pauli paramagnetic gas with indentical spin 1/2 fermions in the presence of an external magnetic field \cite{Ref4}. From the grand canonical distribution through the Fermi-Dirac statistics, the thermodynamic potential can be obtained as follows:
  \begin{equation}
  \phi (x,y,z) = I x^{-{3\over 2}}[f_{5\over 2}(e^{-y-Jz})+ f_{5\over 2}(
e^{-y+Jz})].
\end{equation}    
where $I={{(2\pi m)^{3\over 2}}\over h^3}$ and 
$x=1/T, y = -\mu/T, z= -H/T $ are thermodynamic coordinates.  $T$, $\mu$, and $H$ are also temperature, chemical potential, and external magnetic field, respectively. The $f_n(\eta )$ function is defined by:
\begin{equation}
f_n(\eta ) = {1\over \Gamma (n)} \int_0^\infty {{X^{n-1} dX}\over {e^X\over 
\eta } +1}
\end{equation}
where $\eta = \exp ({\mu \over {K_B T}})$ is called fugacity. The Ruppenier metric in thermodynamic geometry is given by:
\begin{equation}
g_{ij}={\partial^2 \phi \over {\partial Y^i \partial Y^j}}\,\,\ ; \,\,\ Y^i=(x,y,z)
\end{equation}
 The Ricci scalar, ${}^{(3)}R$ was already obtained as a symmetric function of $z$ \cite{Ref4}. It means that the scalar curvature doesn't depend on the orientation of external magnetic field. In the classical limit in the lack of the external magnetic field ($H \to 0$), when $\eta^\pm \rightarrow \eta $ and $f^{\pm}_n(\eta) \rightarrow
\eta$; the ${}^{(3)}R$ is rewritten as follows:
\begin{equation}\label{equation53}
\lim_{H \to 0 (z \to 0)} {}^{(3)}R= -{9 \over{2 \rho}}
\end{equation}
 where in the classical regime, $\rho = {2\over{\lambda^3}} \eta$ and $\lambda = {h / (2 \pi m_0 K_B T)^{1\over 2}}$. Eq. (\ref{equation53}) shows that in the classical limit, the curvature of a Pauli paramagnetic gas depends on the volume occupied by a single particle. The scalar curvature is also similar to the curvature that was obtained for a
two-component ideal gas by Ruppeiner \cite{Ref5}. 
 
In this letter we explore the physical properties of a Pauli paramagnetic gas by studying the intrinsic curvature  of a hypersrface corresponding to a zero magnetic field ( $z \to 0$). The intrinsic curvature of $H$-zero hypersurface can be calculated as:
\begin{equation}
{}^{(2)}R_{in}=-{5 \over{2 \rho}}
\end{equation}
Because the intrinsic curvature is negative, so the statistical interactions of a Pauli paramagnetic gas can be repulsive which indicates a more stable  paramagnetic gas. It should be noted that in the classical limit the extrinsic curvature vanish ($K=0$) and the Gauss-Codazzi relation is held in this case. On the other hand, the Ruppenier curvature for a non-interacting classical paramagnetic gas through Maxwell-Boltzman statistics with the thermodynamic potential, $\phi_c = 4\pi I x^{-{3\over 2}} e^{-y} {{\sinh (Jz)}\over{Jz}}$ ($J$ is magnetic momentum) \cite{Ref4}, takes the following form:
 \begin{equation}
{}^{(3)}R_{c}={\frac {{\lambda}^{3}Jz}{8\pi \,\eta\,\sinh \left( Jz \right) }}
 \end{equation}
 From the equation of state ($PV=NK_{B}T$), one can see the curvature is proportional to the volume occupied by single particle (${}^{(3)}R_{c}=1/2 \rho$) \cite{Ref4}. It is surprising that in the absence of an external magnetic field, the curvature is not zero (${}^{(3)}R_{c}={\frac {{\lambda}^{3}}{8\pi \,\eta}}$), while we would expect the curvature to be zero because of non-interaction particles \cite{Ref1}. To obtain a correct result we have to 
calculate the induced metric on a zero magnetic field hypersurface in the thermodynamic geometry. It can easily be shown that $H$-zero hypersurface is flat which indicates a non-interacting gas as expected. Our analysis indicates
that the intrinsic  geometry of hypersurfaces in thermodynamic geometry have important physical information. In order to have correct information we must study hypersurfaces in thermodynamic geometry.

\section{conclusion}\label{S4}
This work  analyzed the thermodynamic geometry of a black hole from the perspective of an  extrinsic curvature.
It was found that the extrinsic scalar curvature represents the critical behavior  of  a second order phase transition in a thermodynamic system. Some particular Ricci tensor elements were found to have the same sign behavior as heat capacities.
Another part of the article explained the relationship between the intrinsic, the extrinsic, and the total curvatures of thermodynamic geometry of  a  system by sitting on a certain hypersurface. For the KN-AdS black hole on a constant $J$ hypersurface, the curvature scalar was broken down into two parts. One was a zero intrinsic curvature (the Ruppeiner curvature of the RN black hole) while the other was an extrinsic part whose divergence points were the singularities of a non-rotating KN-AdS black hole. We also used the intrinsic curvature of the relevant hypersurface to investigate some thermodynamic properties such as stability and the statistical interaction.

As a result, the critical behavior of a thermodynamic system on an explicit hypersurface can be explained consistently by using intrinsic and  extrinsic curvatures of this hypersurface.

\appendix
\section{Partial derivative and bracket notation}\label{A1}
 When $f$,  $g$, and $h$ are explicit functions of ($a, b$), we can obtain the following relation for the partial derivative.
\be
{{\left( \frac{\partial f}{\partial g} \right)}_{h}}=\frac{{{\left\{ f,h \right\}}_{a,b}}}{{{\left\{ g,h \right\}}_{a,b}}}
\label{N15}
\ee
where,
\be
{{\{f,h\}}_{a,b}}={{\left( \frac{\partial f}{\partial a} \right)}_{b}}{{\left( \frac{\partial h}{\partial b} \right)}_{a}}-{{\left( \frac{\partial f}{\partial b} \right)}_{a}}{{\left( \frac{\partial h}{\partial a} \right)}_{b}}
\ee
Moreover, if one considers $a=a(c,d)$ and $b=b(c,d)$, Eq. (\ref{N15}) can be rewritten as:
\begin{equation}
{{\left( \frac{\partial f}{\partial g} \right)}_{h}}=\frac{{{\left\{ f,h \right\}}_{a,b}}}{{{\left\{ g,h \right\}}_{a,b}}}=\frac{{{\left\{ f,h \right\}}_{c,d}}}{{{\left\{ g,h \right\}}_{c,d}}}
\label{N20}
\end{equation}
And, the determinant of the Jacobian transformation can be written in the bracket notation as in the form below:
\begin{equation}
det \left[\frac{\partial(f,g)}{\partial{(h,k)}} \right]=det\left( \begin{array}{cc}
  \noalign{\medskip}  {{\left( \frac{\partial f}{\partial h} \right)}_{k }} & {{\left( \frac{\partial f}{\partial k} \right)}_{h}}  \\
    \noalign{\medskip} {{\left( \frac{\partial g}{\partial h} \right)}_{k}} & {{\left( \frac{\partial g}{\partial k} \right)}_{h}}  \\
 \end{array} \right)=\frac{{{\left\{ f,g \right\}}_{a,b}}}{{{\left\{ h,k \right\}}_{a,b}}}
\end{equation}
 Generally,  the partial derivative for functions with $n+1$ variables can be calculated as follows:
\begin{equation}
{{\left( \frac{\partial f}{\partial g} \right)}_{{{h}_{1}},.....,{{h}_{n}}}}=\frac{{{\left\{ f,{{h}_{1}},...,{{h}_{n}} \right\}}_{{{q}_{1}},{{q}_{2}},...,{{q}_{n+1}}}}}{{{\left\{ g,{{h}_{1}},...,{{h}_{n}} \right\}}_{{{q}_{1}},{{q}_{2}},...,{{q}_{n+1}}}}}
\end{equation}
where, the $f$, $g$, and ${{h}_{n}}$ ($n=1,\,2,\,3,\,...$) are functions of ${{q}_{i}},i=1,...,n+1$ variables \cite{ref20} and,
\begin{eqnarray}
{{\left\{ f,{{h}_{1}},...,{{h}_{n}} \right\}}_{{{q}_{1}},{{q}_{2}},...,{{q}_{n+1}}}}=\\
 \nonumber \sum\limits_{ijk....l=1}^{n+1}{{{\varepsilon }_{ijk...l}}\frac{\partial f}{\partial {{q}_{i}}}}\frac{\partial {{h}_{1}}}{\partial {{q}_{j}}}\frac{\partial {{h}_{2}}}{\partial {{q}_{k}}}...\frac{\partial {{h}_{n}}}{\partial {{q}_{l}}}
\end{eqnarray}

\section{The concepts of extrinsic and intrinsic curvatures for a hypersurface }\label{A2}
 In this section, we briefly review the concept of extrinsic curvature. For an n-dimensional manifold $\mathfrak{M}$,
a special hypersurface $\Sigma $ can be defined as follows:
\begin{equation}
\Phi ({{x}^{\alpha}})=0
\end{equation}
where, ${{x}^{\alpha}}$s are the coordinates of the manifold $\mathfrak{M}$. The induced metric on $\Sigma $ can be written as:
\begin{equation}
ds_{\Sigma }^{2}={{g}_{\alpha \beta }}d{{x}^{\alpha }}d{{x}^{\beta }}={{g}_{\alpha \beta }}(\frac{\partial {{x}^{\alpha }}}{\partial {{y}^{a}}}d{{y}^{a}})(\frac{\partial {{x}^{\beta }}}{\partial {{y}^{b}}}d{{y}^{b}})={{h}_{ab}}d{{y}^{a}}d{{y}^{b}}
\end{equation}
where,
\begin{equation}
{{h}_{ab}}={{g}_{\alpha \beta }}E_{a}^{\alpha }E_{b}^{\beta } ; E_{a}^{\alpha }=\frac{\partial {{x}^{\alpha }}}{\partial {{y}^{a}}}
\end{equation}
${{h}_{ab}}$ defines the induced metric on the hypersurface. A unit normal ${{n}_{\alpha}}$ can  be introduced if ${{n}_{\alpha}}{{n}^{\alpha}}=\epsilon$ where $\varepsilon =1$ when $\Sigma $ is timelike and $\varepsilon =-1$ when $\Sigma $ is spacelike \cite{ref24}. We select that ${{n}^{\alpha}}$ points in the direction of increasing $\Phi $: ${{n}^{\alpha}}{{\Phi }_{,\alpha}}\rangle 0$. We can also easily show that:
\begin{equation}
{{n}_{\alpha}}=\epsilon \frac{{{\Phi }_{,\alpha}}}{{{\left| {{g}^{\alpha \beta}}{{\Phi }_{,\alpha}}{{\Phi }_{,\beta}} \right|}^{\frac{1}{2}}}}
\end{equation}
 The inverse of the induced metric is obtained as follows:
 \begin{equation}
{{h}^{ab}}E_{a}^{\alpha }E_{b}^{\beta }= {{g}^{\alpha \beta }}-\varepsilon {{n}^{\alpha }}{{n}^{\beta }}
 \end{equation}
One can also introduce the extrinsic curvature tensor on the hypersurface $\Sigma$  using the following relation:
\begin{equation}
{{K}_{a b}}={{n}_{\alpha;\beta}}{E^{\alpha}}_{a}{E^{\beta}}_{b}=\frac{1}{2}{{L}_{n}}{{g}_{\alpha \beta}}{E^{\alpha}}_{a}{E^{\beta}}_{b}
\end{equation}
where, the symbol $;$ and ${{L}_{n}}$ are the covariant  and Lie derivatives of ${{g}_{\alpha \beta}}$ along ${{n}^{\alpha}}$, respectively. Therefore, the extrinsic curvature is defined by:
\begin{equation}\label{ee8}
K={{h}^{a b}}{{K}_{a b}}=n_{;\alpha}^{\alpha}=\frac{1}{\sqrt{ g}}{{\partial }_{\alpha }}\left( \sqrt{ g}{{n}^{\alpha }} \right)
\end{equation}
where, $g=det(g_{\alpha  \beta})$. Suppose that a two-dimensional manifold is embedded in a three-dimensional space. The induced metric ${{h}_{ab}}$ and the extrinsic curvature ${{K}_{ab}}$ contain the necessary information about the properties of the hypersurface $\Sigma$. The full Riemann curvature tensor of the 3-dimensional space and the curvature tensor of the 2-dimensional hypersurface are related by the Gauss-Codazzi equation:
\begin{equation}
{}^{(3)}{{R}_{\alpha \beta \gamma \sigma }}E_{a}^{\alpha }E_{b}^{\beta }E_{c}^{\gamma }E_{d}^{\sigma }={}^{(2)}{{R}_{abcd}}+\varepsilon ({{K}_{ac}}{{K}_{bd}}-{{K}_{ad}}{{K}_{bc}})
\end{equation}
This indicates that the three-dimensional Riemann tensor can be expressed in terms of the intrinsic and extrinsic curvature tensors of the hypersurface \cite{ref27}. Writing the Gauss-Codazzi equation in the contracted form, we obtain:
\begin{equation}\label{ee11}
{}^{(3)}R={}^{(2)}R_{in}+ \epsilon ({{K}^{2}}-{{K}_{a b}}{{K}^{a b}})+2 \epsilon (n_{;\beta}^{\alpha}{{n}^{\beta}}-{{n}^{\alpha}}n_{;\beta}^{\beta})_{;\alpha}
\end{equation}
This relation is the three-dimensional Ricci scalar evaluated on the hypersurface $\Sigma$.
The ${}^{(2)}R$ is intrinsic Ricci scalar of the hypersurface. The third term, on the right hand  side of the Equation, can be expressed as:
\begin{equation}\label{eee1}
{{K}_{a b }}{{K}^{a b }}=n_{;\beta }^{\alpha }n_{;\alpha }^{\beta }
\end{equation}
For the two-dimensional space, the hypersurface $\Xi $ is a one-dimensional space with the normal vector ${{n}_{a}}=\left( 1/\sqrt{\left| {{g}^{ab}} \right|} \right)\delta _{a}^{b}$. For this case, we have:
\begin{equation}
{{K}^{2}}={{K}_{ab}}{{K}^{ab}}
\end{equation}
Thus, the Ricci scalar ${}^{(2)}R$ is determined as follows:
\begin{equation}\label{ee9}
{}^{(2)}R=2\epsilon (n_{;b}^{a}{{n}^{b}}-{{n}^{a}}n_{;b}^{b})_{;a}
\end{equation}




\end{document}